# GV-Index: Scientific Contribution Rating Index That Takes into Account the Growth Degree of Research Area and Variance Values of the Publication Year of Cited Paper

Akira Otsuki [1] and masayoshi Kawamura[2]

[1] Tokyo Institute of Technology, Tokyo, Japan
`cecil2005@hotmail.co.jp`
[2] MK future software, Ibaraki, Japan
`kawamura.masa@nifty.com`

## Abstract

*There are a wide variety of scientific contribution rating indices including the impact factor and h-index. These are used for quantitative analyses on research papers published in the past, and therefore unable to incorporate in the assessment the growth, or deterioration, of the research area: whether the research area of a particular paper is in decline or conversely in a growing trend. Other hand, the use of the conventional rating indices may result in higher rates for papers that are hardly referenced nowadays in other papers although frequently cited in the past. This study proposes a new type of scientific contribution ranking index, "Growing Degree of Research Area and Variance Values Index (GV-Index)". The GV-Index is computed by a principal component analysis based on an estimated value obtained by PageRank Algorithm, which takes into account the growing degree of the research area and its variance. We also propose visualization system of a scientist's network using the GV-Index.*

## 1. Introduction

As typical scientific contribution indexes, such as *h*-Index, *g*-Index, *A*-Index and *R*-Index, have been conventionally assessed based on literatures published in the past, these values tend to be higher in case of well-experienced scientists or those who have larger number of colleagues. In addition, if quoted by many papers in the past, an index value will be highly computed even if these papers have not been cited current. Therefore, this study will calculate "The growing degree of the research area" and "Variance values of the publication year of the cited literature" as an observation value of principal component analysis. Then we propose a method for calculating new synthetic variables (scientific contribution estimated index for scientist) by conducting principal component analysis based on these two observation values in the study.

## 2. Preceding Studies

This section describe about preceding studies at 2.1 later.

## 2.1. h-index

*h*-index is the index Hirsch, J.E. [1] did proposal. One scientist of h-index is satisfies following matter. The number of papers that the number of citations is more than *h* is more than *h*. We show example of *h*-Index in the Table.1.

Table.1 Example of *h*-Index

| Scientist | Papers (Number of citation) | *h*-index |
|---|---|---|
| A | Paper A(9), Paper B(7), Paper C(5), Paper D(4), Paper E(4) | 4 |
| B | Paper A(35), Paper B(9), Paper C(5), Paper D(3), Paper E(1) | 3 |

## 2.2. g-index

Egghe, L. [2] proposed the *g*-index as a modification of the *h*-index. For the calculation of the *g*-index, the same ranking of a publication set -paper in decreasing order of the number of citations received- is used as for the *h*-index. Egghe defines the *g*-index "as the highest number *g* of papers that together received $g^2$ or more citations (1). In contrast to the *h*-index, the *g*-index gives more weight to highly cited papers.

$$g \leq \frac{1}{g} \sum_{i \leq g} c_i$$

(1)

## 2.3. A-index

The proposal to use this average number of citations as a variant of the *h*-index was made by Jin [3]. Jin introduced the *A*-index (as well as the *m*-index, *r*-index, and *AR*-index) includes in the calculation only papers that are in the Hirsch core. It is defined as the average number of citations of papers in the Hirsch core.

$$A = \frac{1}{h} \sum_{j=1}^{h} cit_j$$

(2)

## 2.4. R-index

The better scientist is 'punished' for having a higher *h*-index, as the *A*-index involves a division by h. Therefore, instead of dividing by *h*, the authors suggest taking the square root of the sum of citations in the Hirsch core to calculate the index. Jin et al. [4] did proposal to this new index as the *R*-index, as it is calculated using a square root. R-index- measures the citation intensity in the Hirsch core, the index can be very sensitive to just a very few papers receiving extremely high citation counts (3).

$$R = \sqrt{\sum_{j=1}^{h} cit_j}$$

(3)

## 2.5. Problem so far of preceding studies

Problem so far of scientist evaluation index is these are used for quantitative analyses on research papers published in the past, and therefore unable to incorporate in the assessment the growth, or deterioration, of the research area: whether the research area of a particular paper is in decline or conversely in a growing trend. On the other hand, all indexes above are not considered about a growth rate of research area. Namely, if these research areas have already obsolete meaningless, even if the paper has a lot of citations. We think that it is very important to consider about a growth rate of research area. But the past scientist evaluation indexes is not consider about a growth rate of research area. From mentioned above, the past scientist evaluation indexes has an issues of quality assessment yet. Therefore, we will show the concept of this study by next chapter aiming at improvement progress of quality assessment.

# 3. CONCEPT

In order to solve prior chapter problem, we will calculate using principal component analysis based on two observed following values. This calculated index called "GV-index (Growing degree of research area and Variance values index)". GV-index is intended for journal papers.

① Growing Degree of Research Area
② The Page Rank algorithm considering the degree of dispersion of the cited papers year

① the above is the value to evaluate whether there is a growing trend in the research area. Also ② the above is the value to evaluate the importance of scientists. The PageRank algorithm [8] is a technique used to determine the most "important" page quantitatively by using calculations in the presence of mutual referencing relations such as hyperlink structures. In this study, the strictness of each paper is calculated using this algorithm. That is to say, assuming that the sum of the scores of the citations that "flow out" to each paper and the sum of the scores of the citations that "flow in" from each paper are equal to each other, such a sum is then considered as the score of the pertinent paper, and papers with higher scores are considered more important. By applying the variance value to calculation of the score of citations that "flow in" from each paper, it is possible to identify the key papers in each area. Although scores have been assigned equally in the conventional algorithm when there are multiple citations that "flow in," the severities reflecting the state of variance in the citation year are calculated in this study with the consideration that more citations will "flow in" to papers with higher variance values. We propose a new scientist evaluation index by principal component analysis using this two observation values.

## 3.1. Calculation of Cluster growth

First, calculate the Cluster growth rate as observed values of principal component analysis. The Cluster of this study is based on random network. Random network was proposal by Paul Erdös and Alféd Rényi [9-11] at 1960. The random network is the network that there are random edges in among the nodes. We will use Newman method as the clustering method in this study. Then the group of papers identified by clustering were labelling of research area by experts. We describe the steps to create a random network. Assume the total number of nodes to be "$N$", and the probability of existence of each edge to be "$p$". Also assume, at first, $N$ nodes are prepared. In this case, maximum possible number of edges is shown as underline (4).

$$_N C_2 = \frac{N(N-1)}{2}$$
(4)

Each edge is made with the same probability of $p$. The result, random network based on probability $p$, has as many edges as underline (5) on average.

$$p_N C_2 = p \frac{N(N-1)}{2}$$
(5)

Also, average number of edges per node $<k>$ is shown as underline (6).

$$<k>=p(N-1)$$
(6)

As we've seen, a random network can be made by giving $N$ and $p$. In a random network, each edge between nodes exists at the same probability and is not clustered. In random network, the probability that randomly chosen two nodes are linked together equals to $p$, so the clustering coefficient in random network is shown as underline (7).

$$C_{rand} = p = \frac{\langle k \rangle}{N}$$

(7)

Then, calculate the clustering coefficient for each fiscal year. For example, to calculate clustering coefficient fiscal year by fiscal year from FY2008 through FY2012, the clustering coefficient in FY2008 will be the initial value. This will be called $SC_{rand}$. Then, the year to be "$Y$" and the cluster coefficient by year to be $YC_{rand}$. Next, we will calculate the "Growing Degree of Research Area" by the following equation (8) by applying the GACR (Compound Average Growth Rate) [12].

$$CG^Y = \left[\frac{YC_{rand}}{SC_{rand}}\right]^{\frac{1}{Y-1}} - 1$$

(8)

$1/(Y-1)$ is intended for adjusting the elapsed years. Then we will calculate "$CG^Y$" in each fiscal year to date from the publication year of the paper and we will use $CG^Y$ as observed values of principal component analysis.

## 3.2. Calculation of importance of scientists by the Page Rank algorithm considering the degree of dispersion of the cited papers year

Calculation of importance of scientists by the Page Rank algorithm considering the degree of dispersion of the cited papers year as observed values of principal component analysis. First, investigate the period in which was cited by investigating the variance (standard deviation) of the publication years of the cited papers. In this case, the common method for obtaining the standard deviation is expressed as follows.

$$S^2 = \frac{1}{n}\sum_{i=1}^{n}(\bar{P} - P_i)^2$$

(9)

We method for obtaining the standard deviation is expressed as formula (9). We assume the $P_1$, $P_2 \cdots P_{n-1}$, $P_n$ as a period sample. Then, we regard a $\bar{P}$ is the arithmetic average of these. Then we will calculate variance values formula (9) as an arithmetic mean of $(\bar{P} - P_i)^2$. And the obtained value of standard deviation is stored as variance. Then, apply this variance values to the Page Rank algorism [13]. The formula of the Page Rank algorism is expressed as follows.

$$PR(V_i) = (1-d) + d * \sum_{V_j \in I_n(V_i)} \frac{PR(V_j)}{|Out(V_j)|}$$

(10)

$d$ is the parameter, and can be any real number within the range [0,1]. Formula (10) starts at any value that is given to each node in the graph, and is repeatedly calculated until the value is not exceeding the designated threshold value. Once the calculations are complete, the most important node is determined. Formula (10) is set so that the sum total of the inlink score and the sum total of the outlink score is equal, and as this sum total is seen as the page score, designating pages with higher scores more valuable. However, we method applies the variance value of formula (9) to the score calculations of the inlink and the outlink. While past algorithms would, distribute scores evenly when there were multiple outlinks for example, we method would calculate based on the thought that the points will flow towards higher variance. As a result, importance of scientists can be calculated in a way that reflects the dispersity of the referenced year. This formula is expressed as follow.

$$PR^Y = PR^S(V_i) = (1-d) + d * \sum_{V_j \in In(V_i)} S_i^{\wedge}2 \frac{PR^S(V_j)}{\sum_{V_k \in Out(V_j)} S_k^{\wedge}2}$$

(11)

The $Y$ of $PR^Y$ represents the relevant year. Because the variance values is expressed as a "Standard deviation$^{\wedge}$2" generally, we will express $S_i^{\wedge}2$ as inlink and $S_k^{\wedge}2$ as outlink. Finally, we will calculate "$PR^Y$" in each fiscal year to date from the publication year of the paper and we will use $PR^Y$ as observed values of principal component analysis.

## 3.3. Calculation of Scientific Evaluation Index "GV-index" using Principal Component Analysis

We will calculate the scientific evaluation index using principal component analysis based on the observation value of the previous section. This index is called "GV-index". Principal component analysis is a mathematical procedure that produces a synthesis of a new one variable from two or more variables. The first, we will prepare the data frame of the observation value (*table.2*). Next, we calculate principal component analysis using the data frame of the observation value.

Table.2 The example of the observation value data frame

| No | $CG^Y$ | $PR^Y$ |
|---|---|---|
| 1 | $CG^{2012}$ | $PR^{2012}$ |
| 2 | $CG^{2011}$ | $PR^{2011}$ |
| 3 | $CG^{2010}$ | $PR^{2010}$ |
| 4 | $CG^{2009}$ | $PR^{2009}$ |
| 5 | $CG^{2008}$ | $PR^{2008}$ |

Principal component loadings are the weight of each variable describes the synthesis variables. This is the partial regression coefficient at regression analysis. In accordance with this, the synthesis variables (*GV-index*) express follows (12) in this study.

$$GV\text{-}index = Comp.1 * CG^Y + Comp.1 * PR^Y$$

(12)

$n$ is integer. And *Comp.1* is the first synthesis variable of $CG^Y$ and $PR^Y$. We will use two variables after standardization. Then we will explain effect for GV-index of observations value. If the values of $CG^Y$ and $PR^Y$ are negative minus, GV-index will take the negative effect. Conversely, if the values of $CG^Y$ and $PR^Y$ are positive plus, GV-index will take the positive effect. Therefore GV-index will be able to consider the cluster growth and importance of scientists by Page Rank algorithm considering the degree of dispersion of the cited papers year.

### 3.4. Visualization of "GV-index"

Fig.1 is the image of visualization of the GV-index. The left side is "reference-between clusters (Communities) relation". From the top are listed in descending order of the cluster size. Also to indicate the reference relationship between the clusters, we will shows two same clusters list. Each cluster will show research areas. Then right side is the scientist's network map. If you select one cluster from clusters list in the left side, will show details (Scientist Network Map) of that cluster to right side. Each node is the scientists. And will show large nodes as the GV-index value is greater. This scientist network map shows one's cluster (research area) as well as also another cluster (another research area). We will be able to comprehend scientist's importance or position of scientists in the research area. Furthermore, we will be able to understand the scientists of affecting other research areas.

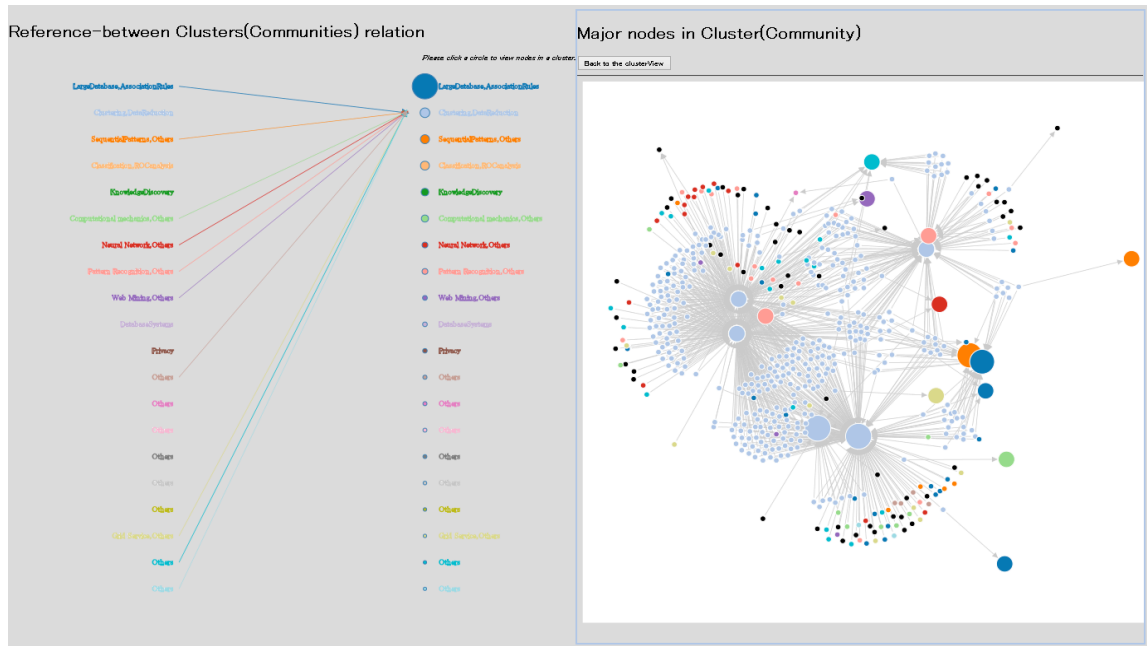

Figure 1. The image of visualization of GV-index (Citation Network Map)

## 4. EVALUATION EXPERIMENT

### 4.1. Outline of Evaluation Experiment and Result of Evaluation Experiment

This section will confirm effectiveness of GV-index by compare h-index, g-index, A-index and R-index. First, we will get the papers from "Web of Science (journal database)" [14] using "Data Mining" query. The search period is 2012 from 1960. Papers numbers we have acquired is about 31,000. Table3 shows the papers of cited number top 5 in 31,000. We decided these four scientists (Table3) as target scientists of this evaluation experiment.

Table.3 The papers of cited number top 5
（Journal DB is the Web of Science, and Query is the "Data Mining"）

|  | Citation | Author's | Paper's name | Publication year |
|---|---|---|---|---|
| 1 | 3649 | **Agrawal,R., et.al** | Mining association rules between sets of items in large databases | 2009 |
| 2 | 1594 | **Fawcett,T., et.al** | An introduction to ROC analysis | 2006 |
| 3 | 1340 | **Zimmermann, P. et.al** | GENEVESTIGATOR. Arabidopsis microarray database and analysis toolbox | 2004 |
| 4 | 1305 | **Agrawal, R., et.al** | Mining sequential patterns | 2005 |
| 5 | 749 | **Foster, I. et.al** | Grid services for distributed system integration | 2002 |

Then, Table4-7 is the papers list about five scientists. These papers lists satisfy the following two conditions. "Four scientists (Table3) are the first author" and further "Research area is the Data Mining". Then Table4-7 are show the result of calculation of $h$-index, $g$-index, $A$-index, $R$-index and GV-index. Calculation method of GV-index will describe in the next section.

Table.4 Compare of each index in Agrawal,R

| GV-index | h-index | g-index | | A-index | R-index | Number of Citation | Name of the Papers |
|---|---|---|---|---|---|---|---|
| 1.353000 | | 1(1) | 3649 | | | 3649 | Mining association rules between sets of items in large databases, 1993. |
| 1.361491 | | 2(4) | 4954 | | | 1305 | Mining sequential patterns, 1995. |
| 1.363350 | | 3(9) | 5688 | | | 734 | Privacy-preserving data mining, 2000. |
| 1.373500 | | 4(16) | 6328 | | | 640 | Automatic subspace clustering of high dimensional data for data mining applications, 1998. |
| 1.292764 | | 5(25) | 6813 | | | 485 | Database mining: A performance perspective, 1993. |
| 1.352239 | | 6(36) | 7211 | | | 398 | Parallel mining of association rules, 1996. |
| 1.373296 | | 7(49) | 7282 | | | 71 | Automatic subspace clustering of high dimensional data, 2005. |
| 1.277639 | **8** | **8**(64) | 7313 | **914.125** | **85.52** | 31 | Securing electronic health records without impeding the flow of information, 2007. |
| **1.343410** | | | | | | | |

Table.5 Compare of each index in Fawcett,T.

| GV-index | h-index | g-index | | A-index | R-index | Number of Citation | Name of the Papers |
|---|---|---|---|---|---|---|---|
| 1.244574 | | 1(1) | 1613 | | | 1613 | An introduction to ROC analysis, 2006. |
| 1.373041 | | 2(4) | 2062 | | | 449 | Adaptive fraud detection, 1995. |
| 1.247522 | | 3(9) | 2112 | | | 50 | Using rule sets to maximize ROC performance, 2001. |
| 1.146067 | **4** | **4**(16) | 2125 | **531.25** | **46.10** | 13 | PRIE: A system for generating rulelists to maximize ROC performance, 2008. |
| **1.252801** | | | | | | | |

Table.6 Compare of each index in Zimmermann, P.

| GV-index | h-index | g-index | | A-index | R-index | Number of Citation | Name of the Papers |
|---|---|---|---|---|---|---|---|
| **1.404585** | **1** | **1**(1) | 1226 | **1226** | **35.01** | 1226 | GENEVESTIGATOR. Arabidopsis microarray database and analysis toolbox |

Table.7 Compare of each index in Foster, I.

| GV-index | h-index | g-index | | A-index | R-index | Number of Citation | Name of the Papers |
|---|---|---|---|---|---|---|---|
| 1.403287 | | 1(1) | 749 | | | 749 | Grid services for distributed system integration, 2002. |
| 1.400020 | **2** | **2**(4) | 782 | **391.00** | **27.96** | 33 | Data integration in a bandwidth-rich world, 2003. |
| **1.401654** | | | | | | | |

## 4.2. Calculation of GV-index

First, will prepare two observation values ($CG^Y$ and $PR^Y$) previously described. Then prepare the data frame like table8. Table8 is the sample of Zimmermann, P.

Table.8 The sample of observation values data frame

| No | $CG^Y$ | $PR^Y$ |
|------|------|------|
| 2012 | 1319 | 0.02269 |
| 2011 | 1205 | 0.02336 |
| 2010 | 1057 | 0.02521 |
| 2009 | 914 | 0.02720 |
| 2008 | 727 | 0.02744 |

This study uses the R-2.15 [15] as a principal component analysis tool. There are two functions for a principal component analysis tool in the R. It's called prcomp() and princomp(). Because do not make much difference, this study uses princomp().fig.2 is the example of the result of the principal component analysis using princomp(). The argument "cor=TRUE" is designate that analysis to standardize the raw data. Other hand, the principal component loadings can be output by specifying the "loadings = TRUE" argument to "summary ()".

```
> data <- read.csv("/mydata.csv",head=F)
> data2 <- princomp(data, cor=TRUE)
> data2
Call:
princomp(x = data, cor = TRUE)
Standard deviations:
```

| Comp.1 | Comp.2 |
|------|------|
| 1.4045851 | 0.1647446 |

```
2 variables and 9 observations.

> summary(data2, loadings=TRUE)
Importance of components:
```

| | Comp.1 | Comp.2 |
|------|------|------|
| Standard deviation | 1.4045851 | 0.16474458 |
| Proportion of Variance | 0.9864296 | 0.01357039 |
| Cumulative Proportion | 0.9864296 | 1.00000000 |

```
Loadings:
```

| | Comp.1 | Comp.2 |
|------|------|------|
| V1 | -0.707 | -0.707 |
| V2 | 0.707 | -0.707 |

```
>
```

Figure 2. Sample of result of the principal component analysis (using R-2.15)

Because principal component analysis will create synthesis variable of the same numbers as the number of observed variables, we will need adopt the synthetic variable of the appropriate number, then It is necessary to truncates other composite variables. This study will use standard deviation and cumulative contribution this evaluation criteria. In the principal component analysis, the weight will calculate so that the variance of synthetic variable maximized. Therefore, we can say that the bigger of these values, the better synthetic variable. Also, in usually, Cumulative contribution rate use the main component of more than 80%. In this case (Fig.2), because Comp.1 has explained more than 98% (0.98) of all data, it is possible to truncate the below Comp.2. Show the principal component scores (Comp.1) in Table 4-7. The last row of GV-index in the table 4-7 show the average if there are the plural papers.

## 4.3. Discussion for the Result of Evaluation Experiment

The numbers of publication papers of Zimmermann, P. and Foster, I. are small as shown in Table4-7. In contrast, The number of publication papers of Agrawal,R and Fawcett,T.

are large. The values of h-index, g-index, A-index and R-index were rise in proportion to number of publication papers. Fig3-7 shows compare each index. h-index, g-index and R-index had small value at Zimmermann, P. Incidentally, though the calculated A-index value of Zimmermann, P. is high, this is because A-index is an index for calculating square roots and since there is only one applicable report by Zimmermann, P., his A-index was calculated simply by using the citation count (1,226) of the said report. According to an expert, Zimmerman, P. invented the e-mail coding software package Pretty Good Privacy [16], and has served as a Fellow at the Stanford Law School's Center for Internet and Society. However, while being a prospective scientist he has produced only a few applicable papers and as such his calculated h-index, g-index, and R-index values were low. Conversely, the value of GV-index was calculated without being influenced by the low number of published papers.

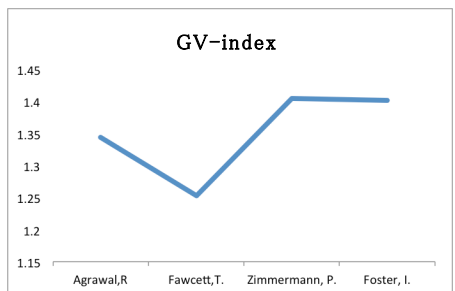
Figure 3. GV-index

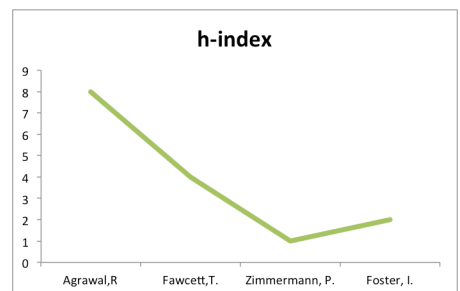
Figure 4. *h*-index

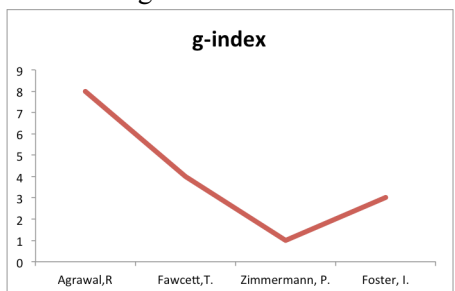
Figure 5. *g*-index

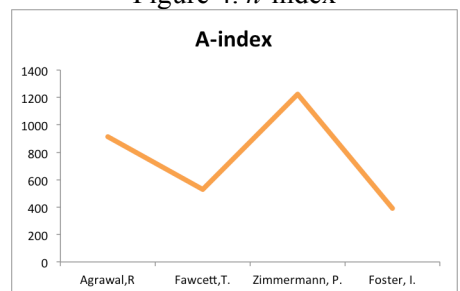
Figure 6. *A*-index

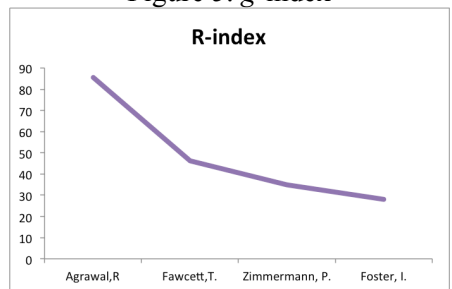
Figure 7. *R*-index

Regarding the previously mentioned experiment results, the causes which enabled a more qualitative evaluation are discussed. Fig.8 shows a comparison of the transition of the cluster size growth rate in Agrawal, R.'s data mining papers (eight in total). In Fig.9, "Mining association rules between sets of items in large databases, 1993. (Hereinafter referred to as Agrawal, R. 1993: Database)" was singularly extracted from Fig.8 to show its cluster size growth rate transition. "Agrawal, R.1993: Database" is the most number of citation (3,649) in this evaluation experiment as shown in Table4. But the number of papers of research area of this paper (Agrawal, R.1993: Database) is declining after 2010 as shown in Fig8. Therefore the value of GV-index was low as shown in Table4.

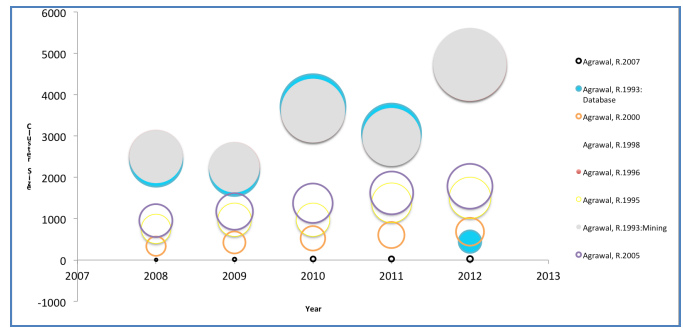

Figure 8. Comparison of the transition of the degrees of growth of the cluster size of the Data Mining articles by Agrawal,R

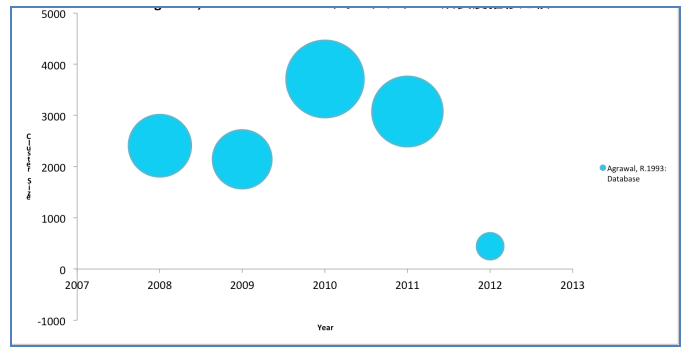

Figure 9. Comparison of the transition of the degrees of growth of the cluster size of the "Agrawal, R.1993:Database"

## 5. CONCLUSION

In this study, we pointed out, having conducted a survey of the past representative scientific contributory indexes that they confined to a narrow sense of the quantitative analyses of the articles published in the past. With respect to this issue, concretely, a problem is explained by the fact that we cannot assess the degree of growth of those research domains concerning whether the research area focused on by an article tends to decline or develop when a research provides only a quantitative analysis. With regard to another problem, it is recognised that even a work which is now not often quoted, albeit it was often cited in the past, can be considered as relevant to the citation counts of the present day scholarship. For the purpose of dealing with these problems, this study, first, calculated the important evaluation value of scientists by the PageRank algorithm as the observation data of principal component analysis, taking account of the degree of growth of the research domain and the variance of the year of the articles quoted, and then, proposed a new scientific contributory index (GV-index) by having carried out principal component analysis on the basis of these data. Further, we did Implementation the scientists network map based on the GV-index. In the result of evaluation experiment, the index values of h-index, g-index, A-index, R-index were larger in proportion to the number of citation. But GV-index could calculation in consideration of the growth rate of the research area.

**Authors**

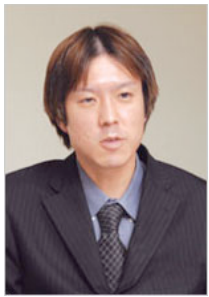

**Akira Otsuki**

Received his Ph.D. in engineering from Keio University (Japan), in 2012. He is currently associate professor at Tokyo institute of technology (Japan) and Officer at Japan society of Information and knowledge (JSIK). His research interests include Analysis of Big Data, Data Mining, Academic Landscape, and new knowledge creation support system. Received his Best paper award 2012 at JSIK. And received his award in Editage Inspired Researcher Grant, in 2012.

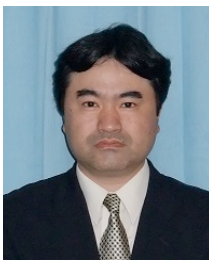

**Masayoshi Kawamura**

Masayoshi Kawamura is a system engineer (Japan). He received M.S. degree from Kyoto Institute of Technology (Japan) in 1998. His research interests include image processing, digital signal processing, and statistical data analysis.